\documentclass[aps,preprint]{revtex4}%
\usepackage{amsfonts}
\usepackage{amsmath}
\usepackage{amssymb}
\usepackage{graphicx}%
\begin{document}
\preprint{ }
\title[Gauge limitations]{Limitations of gauge invariance}
\author{H. R. Reiss}
\affiliation{(Max Born Institute, Berlin, Germany)}
\affiliation{(American University, Washington, DC, USA)}

\begin{abstract}
Although gauge invariance preserves the values of physical observables, a
gauge transformation can introduce important alterations of physical
interpretations. To understand this, it is first shown that a gauge
transformation is not, in general, a unitary transformation. Also, physical
interpretations are based on both kinetic and potential energy expressions.
While the kinetic energy is a measurable quantity, and hence gauge-invariant,
the potential energy is gauge-dependent. Two basic examples are examined; one
classical and the other quantum-mechanical. The aim is to show that the use of
the Coulomb (or radiation) gauge is always consistent with the way that fields
are generated in the laboratory. Upon gauge transformation out of the Coulomb
gauge, this connection is lost, and physical interpretations can give rise to
misleading inferences.

\end{abstract}
\maketitle

\section{Introduction}

There are two principal aims of this paper, although they can be thought of as
two aspects of the same phenomenon. The first aim is to show that physical
interpretations of a specific laboratory scenario are dependent on the gauge
in which the problem is formulated. The second purpose is to demonstrate that
the set of observables that remain invariant under gauge transformations does
not serve to completely characterize the laboratory situation. In other words,
gauge-dependent quantities participate in the description of a physical system.

Gauge invariance is a fundamental principle in both classical and quantum
physics. Its origins can be regarded as being in Newtonian mechanics, where
all physical phenomena arise from applied forces, and forces coming from
electromagnetic phenomena can be expressed directly in terms of the electric
field $\mathbf{E}$ and the magnetic field $\mathbf{B}$. These fields can be
derived from scalar potentials $\phi$ and vector potentials $\mathbf{A}$ that
are not unique. An alteration from one set of potentials to a different set
that define exactly the same fields is called a gauge transformation, and the
requirement that physically measurable quantities should not depend on the
choice of the gauge is called gauge invariance. A complication arises because
the connection between a Newtonian formulation of a physical problem and a
Hamiltonian formulation is not one-to-one. More than one Hamiltonian can
correspond to the same Newtonian equation. This distinction is directly
connected to the fact that Newtonian mechanics is strictly force-dependent,
and hence dependent directly on the fields; whereas Hamiltonians are couched
in terms of potentials, and are therefore gauge-dependent. The most common
form of quantum mechanics is based on the Schr\"{o}dinger equation, so quantum
mechanics is a Hamiltonian formulation and thus directly gauge-dependent in
its mode of expression. A first hint of possible ambiguities comes from the
fact that unitary transformations do not alter physical properties, but a
gauge transformation is generally not a unitary transformation. The lack of
unitarity of a gauge transformation means that what one might call the
\textquotedblleft first line of defense\textquotedblright\ about physical
interpretation is missing.

In the following, after confirming that gauge transformations are not
generally unitary, an elementary classical example of a gauge transformation
is examined. The example is the response of a charged particle to a constant
electric field. This problem is most naturally expressed in terms of a scalar
potential, but it can also be expressed by a vector potential. Both
formulations lead to the same Newtonian equation (as they must), but a
Hamiltonian formulation of the same problem exhibits striking differences in
physical interpretations. For instance, a purely scalar-potential description
is energy-conserving; there is a transfer of energy that is potential into
energy that is kinetic. A vector-potential description does not conserve
energy; there is no potential energy term at all, so that the change in
kinetic energy must be supplied from some source outside the scope of
description of the problem.

The contrast in physical interpretations just mentioned for the constant
electric field example can be ascribed to a departure from the Coulomb gauge
(also known as the radiation gauge) that, for several reasons, is the most
straightforward gauge selection. (See, for example, the textbook by J. D.
Jackson\cite{jackson}.) The Coulomb gauge can be described as that gauge in
which longitudinal fields are represented by scalar potentials and transverse
fields are given in terms of vector potentials. One can be even more specific
about the Coulomb gauge by considering simple limiting cases. Any static
distribution of charges can be described fully by a scalar potential alone.
The other extreme case is that in which no charge or current distributions
whatever exist. The Maxwell equations then allow only one solution: the freely
propagating plane wave. This is a transverse field, where the electric and
magnetic fields are perpendicular to each other and to the propagation
direction. A pure transverse field in the Coulomb gauge is described by a
vector potential alone.

The first example in this paper of the constant electric field illustrates the
advantages of using a scalar potential for a longitudinal field. A physical
origin for a static electric field as it exists between a pair of parallel
capacitor plates is quite natural, whereas a gauge transformation to the use
of a vector potential alters that straightforward view of matters and replaces
it with a situation that seems to be physically unreasonable. Nevertheless,
the classical Newtonian equations of motion are preserved, and gauge
invariance is formally satisfied.

This motivates the selection of the second example considered here, which is
the quantum-mechanical treatment of a free electron in a plane-wave field.
That is, the second example examines a pure transverse field. The known
quantum solution of this problem comes from the Coulomb gauge, and a
representation in any other gauge is found by a gauge transformation from the
Coulomb gauge. That is, a departure from the \textquotedblleft
natural\textquotedblright\ or \textquotedblleft physical\textquotedblright%
\ gauge for description of a transverse field leads to an intractable problem
for solution without recourse to transformation from the Coulomb gauge.

When both longitudinal and transverse fields are simultaneously present, and
physical conditions are appropriate to render the transverse field in terms of
the dipole approximation (i.e., $A\left(  t,\mathbf{r}\right)  \rightarrow
A\left(  t\right)  $ for the vector potential), the Coulomb gauge is commonly
termed the \textit{velocity gauge }in atomic, molecular and optical (AMO)
physics. There then exists a gauge transformation that maintains the
longitudinal field unchanged, but represents the dipole-approximation
transverse field by a scalar potential. This is called the \textit{length
gauge}\ in AMO physics. The length gauge has the practical advantage of
representing all fields by scalar potentials. This is not only mathematically
convenient, but it suggests simple physical interpretations that follow from
the visual superposition of two scalar potentials. This has been done also
with nonperturbatively strong laser fields as long as the wave length is long
enough to justify the dipole approximation and the field is not so strong that
magnetic effects appear.

Nevertheless, there are hazards associated with the departure from the
\textquotedblleft physical gauge\textquotedblright; that is, the Coulomb gauge
(or velocity gauge in dipole-approximation parlance). A physical
interpretation has gained wide currency in the strong-field physics community
that derives from the tunneling point of view that is such a convenient way to
treat an all-scalar-potential AMO problem. Qualitative contradictions seem to
arise that can be traced to the development of an intuition that is not based
on the Coulomb gauge. Specifically, it is customary to describe the relatively
weak-field end of the nonperturbative environment as the \textquotedblleft
multiphoton domain\textquotedblright, because photoelectron spectra reveal
individual peaks that can be associated with specific multiphoton orders. The
stronger-field environment is called the \textquotedblleft tunneling
domain\textquotedblright\ because photoelectron spectra become smooth and
without evidence of specific multiphoton orders. The standard physical picture
in the length gauge envisions a scalar-potential Coulomb attractive well
representing the atomic Coulomb attraction for the electron, being deformed by
a slowly-oscillating linear potential that is used to represent the external
plane-wave field within the length gauge. In such a picture, a bound atomic
electron confronts a finite potential barrier that can be penetrated in the
quantum sense on that side of the Coulomb well that is depressed by the scalar
potential of the laser (i.e. plane-wave) field. In such a picture the
relatively low-intensity nonperturbative domain requires a tunneling process
in order for ionization to occur, whereas the more intense field allows the
initially bound electron the escape over the barrier altogether, with no
tunneling required. This is exactly the reverse of the multiphoton\ domain vs
tunneling domain assignment of names\cite{hr07b}.

Other length-gauge conceptual problems that are not supported by experiments
include the prediction that all photoelectron spectra with linearly polarized
fields are maximal at zero energy, as are all longitudinal momentum distributions.

In the Coulomb gauge, in which the laser field is described by a vector
potential, there is no tunneling concept of ionization, and the conflict
described above does not exist. If an attempt is made to find a putative
\textquotedblleft tunneling limit\textquotedblright\ within the Coulomb gauge
by seeking that part of the ionization transition amplitude that exhibits the
algebraic $\exp\left(  -C/E\right)  $ behavior characteristic of tunneling
(where $E$ is the amplitude of the electric field $\mathbf{E}$), this turns
out to be a small and unrepresentative part of the full ionization amplitude.

Other nonphysical consequences of using a physical intuition that is not
formed within the Coulomb gauge are discussed later.

One further introductory matter must be mentioned. When the total Hamiltonian
$H$ possesses explicit time dependence, then it is not justifiable to regard
the Hamiltonian as the sum of kinetic and potential energies. The kinetic
energy $T$, as a measurable quantity, presents no problems, but the remainder
of the Hamiltonian does. However, the quantity $H-T$ will be referred to
simply as the potential energy. This is common practice in quantum mechanics.

\section{Nonunitarity of Gauge Transformations}

The discussion here is in terms of the Schr\"{o}dinger equation, although
analogous results apply as well in other formulations of quantum mechanics.

For any wave function that describes a charged particle subjected to an
electromagnetic field, a gauge transformation acts through a unitary operator
$U$ to produce the transformed wave function
\begin{equation}
\Psi^{\prime}=U\Psi. \label{a}%
\end{equation}
This is often taken to be sufficient evidence that a gauge transformation is
unitary. However, whereas a unitary transformation changes any operator $O$
according to the rule%
\begin{equation}
O^{\prime}=UOU^{-1}, \label{b}%
\end{equation}
it is easily verified that a gauge transformation of the Schr\"{o}dinger
Hamiltonian operator $H$ produces the gauge-transformed Hamiltonian
$H^{\prime}$ that obeys the rule%
\begin{equation}
H^{\prime}-i\partial_{t}=U\left(  H-i\partial_{t}\right)  U^{-1}, \label{c}%
\end{equation}
where atomic units are used. For direct comparison with Eq. (\ref{b}), the
gauge transformed Hamiltonian can be written as%
\begin{equation}
H^{\prime}=UHU^{-1}-Ui\partial_{t}U^{-1}+i\partial_{t}. \label{d}%
\end{equation}
The contrast between Eqs. (\ref{b}) and (\ref{d}) shows that any
transformation that contains time dependence is automatically nonunitary. This
is decidedly non-trivial, since any gauge change that is limited to being
independent of time cannot alter the scalar potential in any way. In
particular, the transformation between the velocity and length gauges would be excluded.

(Although it is outside the scope of the present article, it must be remarked
that there is a widespread belief in the AMO community not only that a gauge
transformation is a unitary transformation, but that any transition matrix
element is automatically gauge invariant. Both beliefs are unfounded.)

\section{Gauge Transformation in a Classical Longitudinal Field Example}

The purpose of this Section is to show how a simple physical system can be
altered into a seemingly unphysical problem by a gauge transformation. The
example employed here involves a longitudinal field. A pure transverse field
will be considered in Section V.

A particle of charge $q$ in a constant electric field $\mathbf{E}_{0}$ is a
one-dimensional problem since the only direction of consequence is the
direction of $\mathbf{E}_{0}$, taken to define the coordinate axis $x$. The
electric field can be described by the scalar potential%
\begin{equation}
\phi=-E_{0}x, \label{a1}%
\end{equation}
and there is no need for a vector potential:%
\begin{equation}
\mathbf{A}=0. \label{b1}%
\end{equation}
The Hamiltonian of this classical system is%
\begin{equation}
H=p^{2}/2m-qE_{0}x. \label{c1}%
\end{equation}
Hamilton's equations%
\begin{equation}
\overset{\cdot}{x}=\frac{\partial H}{\partial p}=\frac{p}{m},\qquad
\overset{\cdot}{p}=-\frac{\partial H}{\partial x}=qE_{0}, \label{d1}%
\end{equation}
can be combined to give the Newtonian equation%
\begin{equation}
m\overset{\cdot\cdot}{x}=qE_{0}, \label{e}%
\end{equation}
that has the solution%
\begin{equation}
x=\frac{qE_{0}}{2m}t^{2}+\overset{\cdot}{x}\left(  0\right)  t+x(0), \label{f}%
\end{equation}
where the dot over a quantity designates a time derivative, and $\overset
{\cdot}{x}\left(  0\right)  $ and $x(0)$ are the initial velocity and position
of the charged particle.

The obvious physical interpretation of the problem described by Eqs.
(\ref{a1}) - (\ref{f}) is that a particle of mass $m$ and charge $q$ is
released between two parallel capacitor plates with a potential difference
between them.

Now a gauge transformation is introduced. If $\Lambda$ is the generating
function of a gauge transformation, the general expressions for the
transformed potentials are%
\begin{align}
\phi^{\prime}  &  =\phi-\frac{1}{c}\frac{\partial\Lambda}{\partial t}%
\Lambda,\label{g}\\
\mathbf{A}^{\prime}  &  =\mathbf{A}+\mathbf{\nabla.} \label{h}%
\end{align}
When the generating function is%
\begin{equation}
\Lambda=-cE_{0}xt, \label{i}%
\end{equation}
then the initial potentials (\ref{a1}) and (\ref{b1}) are transformed to%
\begin{align}
\phi^{\prime}  &  =0,\label{j}\\
\mathbf{A}^{\prime}  &  =-\mathbf{e}_{x}cE_{0}t, \label{k}%
\end{align}
where $\mathbf{e}_{x}$ is the unit vector in the $x$ direction. The new
Hamiltonian is%
\begin{equation}
H^{\prime}=\frac{1}{2m}\left(  p^{\prime}+qE_{0}t\right)  ^{2}, \label{l}%
\end{equation}
leading to the equations of motion%
\begin{equation}
\overset{\cdot}{x}=\frac{\partial H^{\prime}}{\partial p}=\frac{1}{m}\left(
p^{\prime}+qE_{0}t\right)  ,\qquad\overset{\cdot}{p}{}^{\prime}=-\frac
{\partial H^{\prime}}{\partial x}=0. \label{n}%
\end{equation}
These equations combine to give%
\begin{equation}
m\overset{\cdot\cdot}{x}=qE_{0}, \label{o}%
\end{equation}
which is identical to Eq. (\ref{e}). This verifies gauge invariance in this
simple classical example.

Despite the common solution in the two gauges, the problems described are
totally different when a qualitative description is sought. In the original
gauge, one can envision the laboratory environment as arising from parallel
capacitor plates holding different total charges. The transformed gauge has no
obvious laboratory interpretation. Furthermore, the original gauge has a
time-independent Hamiltonian, associated with energy conservation, whereas the
transformed gauge is explicitly time-dependent, signifying the absence of
energy conservation.

It is instructive to evaluate the kinetic and potential energies that arise
from the common solution (\ref{f}). In the original gauge, the kinetic energy
is simply%
\begin{equation}
T=p^{2}/2m. \label{p}%
\end{equation}
From Eq. (\ref{d1}), $p=m\overset{\cdot}{x}$. Using the solution (\ref{f}) to
find $\overset{\cdot}{x}$ leads to the result%
\begin{equation}
T=\frac{1}{2m}\left[  qE_{0}t+m\overset{\cdot}{x}\left(  0\right)  \right]
^{2}. \label{q}%
\end{equation}
The potential energy $U=-qE_{0}x$ can be written as%
\begin{equation}
U=-T+\frac{m}{2}\left[  \overset{\cdot}{x}\left(  0\right)  \right]
^{2}-qE_{0}x\left(  0\right)  , \label{r}%
\end{equation}
so that the sum of the kinetic and potential energies is%
\begin{equation}
T+U=\frac{m}{2}\left[  \overset{\cdot}{x}\left(  0\right)  \right]
^{2}-qE_{0}x\left(  0\right)  =const. \label{s}%
\end{equation}
Equation (\ref{s}) confirms the conservation of energy in the original gauge.

In the transformed gauge, Eq. (\ref{n}) gives the canonical momentum%
\begin{equation}
p^{\prime}=m\overset{\cdot}{x}-qE_{0}t, \label{t}%
\end{equation}
so that the gauge-transformed kinetic energy is, from Eq. (\ref{n}),%
\begin{align}
T^{\prime}  &  =\frac{1}{2m}\left(  m\overset{\cdot}{x}\right)  ^{2}%
\label{u}\\
&  =\frac{1}{2m}\left[  qE_{0}t+m\overset{\cdot}{x}\left(  0\right)  \right]
^{2}.\label{v}\\
&  =T. \label{w}%
\end{align}
This result is to be expected. The kinetic energy is a measurable quantity,
and so it must be preserved in a gauge transformation. On the other hand, the
potential energy is altered completely. In the original gauge, $T+U=const.$,
as shown in Eq. (\ref{s}). In the transformed gauge, there is no potential
energy term at all, so that%
\begin{equation}
T^{\prime}+U^{\prime}=T^{\prime}\left(  t\right)  . \label{x}%
\end{equation}
This not only differs from Eq. (\ref{s}), but the total energy is explicitly a
function of time.

This elementary example -- a constant electric field -- thus illustrates the
essential points to be made in this paper. The conceptual existence of a
constant electric field existing between capacitor plates has been transformed
into something else that seems not to have a physical interpretation attached
to it. The original gauge is the \textit{laboratory gauge} or the
\textit{physical gauge}. There must exist a charge distribution that can give
rise to a constant electric field.

The fact that a simple system in which energy is conserved can be transformed
to another system in which energy is not conserved is a major qualitative
difference. This happens despite the fact that the gauge-invariant electric
field and kinetic energy are explicitly conserved in the gauge transformation.
Both the electric field and kinetic energy are measurable quantities. It is
the non-gauge-invariant potential energy that is the source of the radical
alteration in the apparent physical context of the problem.

It is worth repeating the conclusion just reached: \emph{even in the presence
of strict gauge invariance, the physical interpretation of the problem has
been changed}.

This simple example illustrates the general principle that the relationship
between the Hamiltonian formulation of mechanics and that of Newton is not
one-to-one. Hamilton's equations in (\ref{d1}) give rise to the Newtonian Eq.
(\ref{e}). The very different Hamilton's equations in (\ref{n}) correspond to
the same Newtonian equation \ref{e}. The fact that the electric and magnetic
fields are preserved in the gauge transformation does not make the two gauges equivalent.

All this stems from the fact that non-Newtonian formalisms (e.g., Hamilton's
equations) imply Newton's equations, but the reverse is not true. Since
quantum physics is usually founded on a Hamiltonian or Lagrangian basis, a
change in gauge has physical implications despite the gauge invariance of
electromagnetic fields. This will be shown in Section V below.

\section{Assessment to This Point}

The elementary problem just considered -- a charged particle in a constant
electric field -- is sufficient to establish the two basic aims of this paper.
The physical interpretation of the laboratory scenario is radically altered by
the change of gauge. What started as the description of a charged particle
responding to a static difference of charge on two capacitor plates, becomes
transformed into something difficult to describe in simple terms. In the
laboratory gauge (i.e., the Coulomb gauge), there is strict conservation of
total energy. The potential energy inherent in the charged particle starting
at the initial conditions, becomes converted progressively to kinetic energy,
with the total of both energy forms remaining constant. After gauge
transformation, the increase of the kinetic energy with time is supplied by
some unspecified external source. Total energy is not conserved because the
external source is somehow able to transfer energy into the system to supply
the kinetic energy that is required by the dynamics of the process.

The above paragraph describes the two very different physical interpretations
that attach to the two different gauges. This is a clear example of the first
goal of this paper: to show that physical interpretations depend on the gauge.
At the same time, the second goal is demonstrated: it is not enough to
preserve the values of all physical observables. In this classical example,
the particle trajectory, that may be written as $\mathbf{r}\left(  t\right)
$, constitutes the observable quantity that is sufficient to define the
behavior of the system. The velocity $\mathbf{v}\left(  t\right)
=\overset{\cdot}{\mathbf{r}}\left(  t\right)  $ is an observable quantity, as
is the kinetic energy $T=m\mathbf{v}^{2}/2$. However, the discussion just
presented of the radical differences in physical interpretations had recourse
to an examination of the potential energy. This cannot be over-emphasized.
\emph{The potential energy plays a crucial role in the physical interpretation
of the physical system, but the potential energy is not a gauge-invariant
quantity}\textit{.}

Another way to view the longitudinal field problem is to note that a pure
longitudinal field can be treated entirely in terms of electrostatics. This is
a complete subject in itself that requires only a scalar potential to describe
it. In that sense, a gauge transformation to a representation of the field by
a vector potential is an alien concept. Although the values of all physical
observables are preserved, a simple physical interpretation is altered to
something entirely different that is no longer simple and natural.

The primary aims of this paper are thus already established. The next section
describes a different consequence of using a gauge that is not the physical gauge.

\section{Gauge Transformation in Quantum Transverse-Field Examples}

As already mentioned, the freely propagating plane wave is the unique solution
of the free-space Maxwell equations when there are no electrical charges or
currents. This is the laboratory environment in which the Coulomb gauge comes
into play in its pure form for a transverse field.

Two examples will be presented. It is sufficient here to examine these
problems in a dipole-approximation context. One example relates to the
solution for a free electron in the presence of a plane-wave field. It is
found that the solution is straightforward in the velocity (Coulomb) gauge,
but is beset by peculiarities in the length gauge. The other example refers to
the ionization of an atomic electron by a plane-wave field. In this second
example, treatment of the problem in the length gauge gives useful results
within its domain of applicability, but suggests completely inappropriate
extensions outside that region.

\subsection{Free charged particle in a plane-wave field}

In isolation, the only relevant solution for a free electron in a plane wave
is the relativistic solution. This is known as the Volkov, or Gordon-Volkov
solution\cite{gordon,volkov}, that can be stated exactly for any wave packet
of unidirectional plane waves. However, it will be convenient here to adopt a
simplification that is widely used. If the free-electron solution is to be
employed within a matrix element that also contains a wave function bounded in
space, then it is usually admissible to treat the free particle in the dipole
approximation. The limitations are that the wavelength of the field should not
be so short as to make the dipole approximation inapplicable, nor should the
field be so strong or of such low frequency as to induce magnetic field effects.

The velocity-gauge Schr\"{o}dinger equation for the \textquotedblleft
dipole-approximation free electron\textquotedblright\ is%
\begin{equation}
i\partial_{t}\Psi^{V}=\frac{1}{2}\left(  \widehat{\mathbf{p}}+\frac{1}%
{c}\mathbf{A}\left(  t\right)  \right)  ^{2}\Psi^{V}, \label{aa}%
\end{equation}
where atomic units are used, electromagnetic quantities like the vector
potential $\mathbf{A}\left(  t\right)  $ are in Gaussian units, the
\textquotedblleft hat\textquotedblright\ symbol over the canonical momentum
signifies the quantum operator $\widehat{\mathbf{p}}=-i\mathbf{\nabla}$, and
the superscript $V$ on the wave function identifies it as being in the
velocity-gauge. The solution of Eq. (\ref{aa}) is simply%
\begin{equation}
\Psi^{V}\left(  \mathbf{r},t\right)  =C\exp\left[  i\mathbf{p\cdot r}-\frac
{i}{2}\int_{-\infty}^{t}d\tau\left(  \mathbf{p}+\frac{1}{c}\mathbf{A}\left(
\tau\right)  \right)  ^{2}\right]  , \label{ab}%
\end{equation}
where $\mathbf{p}$ is the eigenvalue of the canonical momentum operator. It
can also be identified as the kinetic momentum outside the bounds of the
electromagnetic pulse.

In the length gauge, indicated by the superscript $L$, the Schr\"{o}dinger
equation corresponding to Eq. (\ref{aa}) is%
\begin{equation}
i\partial_{t}\Psi^{L}=\left(  \frac{1}{2}\widehat{\mathbf{p}}^{2}%
+\mathbf{r\cdot E}\left(  t\right)  \right)  \Psi^{L}. \label{ac}%
\end{equation}
Equation (\ref{ac}) appears to be more simple than Eq. (\ref{aa}), yet the
solution is normally found by a gauge transformation of the velocity gauge
solution (\ref{ab}). That is, the length-gauge solution is%
\begin{equation}
\Psi^{L}\left(  \mathbf{r},t\right)  =C\exp\left[  i\frac{1}{c}\mathbf{r\cdot
A}\left(  t\right)  \right]  \exp\left[  i\mathbf{p\cdot r}-\frac{i}{2}%
\int_{-\infty}^{t}d\tau\left(  \mathbf{p}+\frac{1}{c}\mathbf{A}\left(
\tau\right)  \right)  ^{2}\right]  . \label{ad}%
\end{equation}
Not only is this result more complicated than Eq. (\ref{ab}), it is written in
terms of the velocity-gauge $\mathbf{A}\left(  t\right)  $ rather than the
length-gauge $\phi=-\mathbf{r\cdot E}\left(  t\right)  $. One way to introduce
the length-gauge potential in place of $\mathbf{A}\left(  t\right)  $ in Eq.
(\ref{ad}) is to invert the velocity-gauge relation%
\begin{equation}
\mathbf{E}\left(  t\right)  =-\frac{1}{c}\partial_{t}\mathbf{A}\left(
t\right)  \label{ae}%
\end{equation}
to obtain%
\begin{equation}
\mathbf{A}\left(  t\right)  =-c\int_{-\infty}^{t}d\tau\mathbf{E}\left(
\tau\right)  . \label{af}%
\end{equation}

All of this maneuvering does not really solve the problem. The new expression%
\begin{equation}
\Psi^{L}\left(  \mathbf{r},t\right)  =C\exp\left[  i\mathbf{p\cdot r}%
-i\int_{-\infty}^{t}d\tau\mathbf{r\cdot E}\left(  \tau\right)  -\frac{i}%
{2}\int_{-\infty}^{t}d\tau\left(  \mathbf{p}-\int_{-\infty}^{\tau}dt^{\prime
}\mathbf{E}\left(  t^{\prime}\right)  \right)  ^{2}\right]  \label{ag}%
\end{equation}
no longer contains the vector potential, but the entire expression is far more
implicit than is Eq. (\ref{ab}) in view of the integrations submerged within
Eq. (\ref{ag}). In contrast with the velocity gauge solution (\ref{ab}), which
can be written expressly in terms of the vector potential $\mathbf{A}\left(
t\right)  $, it is not possible to write the length-gauge solution entirely in
terms of the scalar potential $\mathbf{r\cdot E}\left(  t\right)  $. Further,
in order to achieve this length-gauge result, it was necessary to employ Eq.
(\ref{ae}), which is a velocity-gauge expression. In sum, the velocity-gauge
wave function is easily found in a compact, comprehensible form; whereas the
length-gauge solution is complicated, unwieldy, and far from intuitive. All of
these differences can be ascribed to the fact that the velocity gauge is the
physical gauge for a transverse field, and the length gauge is not.

\subsection{Atomic ionization by a plane-wave field}

A velocity-gauge treatment of the ionization of a single-electron atom by a
plane-wave field employs the Coulomb gauge: the Coulomb attraction between the
atomic electron and the nucleus is represented by the scalar Coulomb
potential, and the action of the plane-wave field is given by a vector potential.

The advantages of a length-gauge treatment of the problem are nevertheless
appealing. If the dipole approximation is applicable, a gauge transformation
that allows both interactions to be expressed by scalar potentials makes
possible elementary addition of the two potentials to provide a picture within
which the plane-wave field appears to be a simple slowly-oscillating electric
field. There is then a view of the interaction in which the Coulomb attractive
potential well is depressed on one side by the electric field, so as to make
possible an escape of the bound electron by tunneling through a finite
barrier, or by escaping over the depressed barrier. This picture can be
applied (within limitations) even when the external field is strong enough to
require nonperturbative treatment. Such a nonperturbative theory is a
tunneling theory\cite{kel,nr,ppt,adk}, and it has been found to be very useful.

A clear problem arises when the non-laboratory gauge (the length gauge) is
used as a guide to extend a problem beyond the domain where the two gauges are
demonstrably gauge-equivalent. The tunneling theory is a length-gauge view of
strong-field ionization that represents by scalar potentials both the Coulomb
potential well of the atomic nucleus and the external laser field. A simple
physical picture then exists that envisions the sum of the two potentials that
has the laser field forcing down alternatively one or the other side of the
Coulomb well. This then presents a limited-width potential barrier that can
allow a bound electron to escape from an atom by tunneling through this
barrier. A nonperturbative tunneling theory of ionization contains only one
scaling parameter, known as the Keldysh gamma parameter%
\begin{equation}
\gamma=\sqrt{\frac{E_{B}}{2U_{p}}}, \label{ah}%
\end{equation}
where $E_{B}$ is the field-free binding energy of the atomic electron in the
atom, and $U_{p}$ is the ponderomotive energy of a free electron in the
external laser field. If $U_{p}$ is given in terms of field intensity $I$ and
field frequency $\omega,$ then%
\begin{equation}
U_{p}=I/\left(  2\omega\right)  ^{2} \label{ai}%
\end{equation}
in atomic units for a linearly polarized field. Substitution of Eq. (\ref{ai})
into (\ref{ah}) yields the Keldysh parameter in\ the form%
\begin{equation}
\gamma=\frac{\omega}{I^{1/2}}\sqrt{2E_{B}}. \label{aj}%
\end{equation}
This has suggested to many investigators that the $\omega\rightarrow0$ limit
can be examined by a tunneling theory with $\gamma\rightarrow0$, and this can
then be regarded as a continuous way to apply a nonperturbative theory to
examine atomic ionization in the classical limit. See Ref.\cite{popov} for a
representative view of a widely held attitude.

The fundamental problem here is that for a laser field, which is a transverse
field, the limit $\gamma\rightarrow0$ for constant intensity $I$ enters a
domain where the magnetic field becomes of importance equal to the electric
field, and the dipole approximation fails. That is, the length-gauge
approximation for a strong plane-wave field breaks down completely, and the
length-gauge theory no longer is capable of describing a plane wave.

This basic problem can be viewed in another way. If $\omega$ is held constant,
and the intensity $I$ is allowed to go to $\infty$, this also corresponds to
$\gamma\rightarrow0.$ By the universality of the $\gamma$ parameter, this
should predict a classical limit that is equivalent to the fixed-$I$,
$\omega\rightarrow0$ case. However, $I\rightarrow\infty$ for a laser field is
plainly a relativistic limit for atomic ionization, the dipole approximation
certainly does not apply, and a transverse field cannot be represented by a
scalar potential.

The non-laboratory (or non-physical) length gauge then predicts inapplicable
and misleading behavior if one uses it to predict phenomena that are outside
the (very limited) domain of applicability of the velocity-gauge to
length-gauge transformation.

\section{Conclusions}

It has been shown that a static-field environment (i.e., a longitudinal-field
environment) produces physical interpretations that are entirely consistent
with the laboratory means of generating a longitudinal field. A gauge
transformation to a velocity-gauge representation preserves the values of all
physically measurable quantities, but introduces physical interpretations that
are foreign to the static-electric-field situation. This proves both that
gauge transformations introduce changes in physical interpretations, and that
physical interpretations not associated with the use of a Coulomb gauge will
give unreasonable physical interpretations that have no conceptual or
inferential value.

In the transverse-field situation that exists in the presence of a laser
field, there also results a physically misleading situation if a gauge
transformation is introduced that describes the laser field as a longitudinal field.

The conclusion that is thereby reached is that the radiation gauge (also known
as the Coulomb gauge) should be employed if physical interpretations are
sought that are consistent with the origins of the fields in the problem. The
use of a gauge that does not represent longitudinal fields by scalar
potentials and transverse fields by vector potentials can give radically
misleading physical interpretations despite the fact that gauge invariance
guarantees the preservation of all measurable quantities.

\section{Coda}

Further developments related to this article may be found in Ref.
\cite{hrjmo}. Among these developments is a demonstration that potentials are
more fundamental than fields; that is, a given configuration of fields does
not always uniquely identify an electromagnetic environment.

\end{document}